\documentclass[a4paper]{VisionStyle}
\usepackage{epsfig}

\begin{document}

\title{Neutron star/supernova remnant associations: the view from Tbilisi}

\author{V.V.\,Gvaramadze\inst{1,2} }

\institute{
   E.K.Kharadze Abastumani Astrophysical Observatory, Georgian Academy of
   Sciences, A.Kazbegi ave. 2-a, Tbilisi\thanks{``Tbilisi is an attractive
   city and a major cultural and educational centre" ({\it The New
   Encyclop{\ae}dia Britannica}, 1998, v. 11, p. 592). For one more
   example of attractive and nice cities see Frisch (2001).} 380060,
   Georgia
\and
   Sternberg State Astronomical Institute, Moscow State University,
   Universitetskij Prospect 13, Moscow 119899, Russia}

\maketitle

\begin{abstract}

We propose a new approach for studying the neutron star/supernova
remnant associations, based on the idea that the supernova
remnants (SNRs) can be products of an off-centered supernova (SN)
explosion in a preexisting bubble created by the wind of a {\it moving}
massive star. A cavity SN explosion of a moving star results
in a considerable offset of the neutron star (NS) birth-place from the
geometrical center of the SNR. Therefore: a) the high transverse
velocities inferred for a number of NSs
through their association with SNRs
can be reduced; b) the proper motion vector of a NS
should not necessarily point away from the geometrical center of
the associated SNR. Taking into account these two facts allow us
to enlarge the circle of possible NS/SNR associations, and could
significantly affect the results of previous studies of
associations. The possibilities of our approach are illustrated with
some examples.
We also show that the concept of an off-centered cavity SN explosion could
be used to explain the peculiar structures of a number of SNRs and for
searches for stellar remnants possibly associated with them.

\keywords{Stars: neutron -- ISM: bubbles -- ISM: supernova remnants}
\end{abstract}

\section{Introduction}

Usually the evaluation of reliability of claimed NS/SNR associations is
based on the use of five criteria formulated by
\cite*{vgvaramadze-B2:kas96}, which come to the following questions:\\
\hskip 5mm
-- do independent distance estimates agree?\\
\hskip 5mm
-- do independent age estimates agree?\\
\hskip 5mm
-- is the implied transverse velocity reasonable?\\
\hskip 5mm
-- is there evidense for any interaction between the NS and SNR?\\
\hskip 5mm
-- does the proper motion vector of the NS point away from the SNR center?\\
The last question is considered the most important one since
``a proper motion measurement has the potential to disprove an
association regardless of the answers to the other questions"
(\cite{vgvaramadze-B2:kas96}).

Sometimes a claimed NS/SNR association is considered as false on the basis
of statistical studies of associations (e.g. \cite{vgvaramadze-B2:gae95},
\cite{vgvaramadze-B2:lor98}). For example, one of the
arguments
against the association of PSR \object{B\,1706-44} with the SNR
\object{G\,343.1-2.3} (\cite{vgvaramadze-B2:nic96}) is based
on the suggestion by \cite*{vgvaramadze-B2:gae95} that young ($< 25\,000$
yr) NSs cannot overrun their parent SNR shells.

However, these approaches neglect two very important effects: the
modification of the ambient medium by the ionizing emission and
stellar wind of massive stars (the progenitors of most of SNe),
and the proper motion of SN progenitor stars. The first effect is
important since it is the subsequent interaction of SN blast waves
with their processed ambient medium (a system of cavities and
shells) that results in the observed SNRs: their structure and
evolution are already known to deviate significantly from those
derived from standard models of SNRs based on the Sedov-Taylor
solution (e.g. \cite{vgvaramadze-B2:shu85};
\cite{vgvaramadze-B2:cio89}; \cite{vgvaramadze-B2:che89};
\cite{vgvaramadze-B2:fra91}). The stellar proper motion should be
considered since it could result (\cite{vgvaramadze-B2:gva02} and
references therein) in a considerable offset of the SN explosion
site from the center of the wind-driven bubble (i.e. from the
geometrical center of the future SNR).

Taking into account these two effects could significantly
affect the results of previous studies of NS/SNR associations, and
allow us to enlarge the circle of possible NS/SNR associations and to
search for new associations.

\section{Reliability of NS/SNR associations}

We now discuss the criteria for evaluating the reliability of
NS/SNR associations proposed by \cite*{vgvaramadze-B2:kas96}.
It is obvious that
the first two criteria should be undoubtedly fulfilled for any
proposed association. But the application of the third and
fifth ones for evaluating of proposed associations is not so
straightforward, since they are based on the assumption that
the SN explosion site coincides with the geometrical center of the
SNR. This assumption, however, could be erroneous in the case of a
density-stratified interstellar medium (e.g. \cite{vgvaramadze-B2:gul74})
or in the case of an off-centered cavity SN explosion
(\cite{vgvaramadze-B2:gva02} and references therein).
In both cases the velocity estimates implied by angular
displacements of NSs from the geometrical centers of associated
SNRs could be significantly in error, while the associations
rejected on the basis of high implied transverse velocities of NSs
[e.g. \object{SGR\,0525-66}/SNR \object{N\,49}
(\cite{vgvaramadze-B2:kas00}) or PSR \object{B\,1706-44}/SNR
\object{G\,343.1-2.3} (\cite{vgvaramadze-B2:nic96})] could be genuine.

It is clear that the proper motion vector of a NS born in an
off-centered cavity SN explosion could be oriented arbitrarily with respect
to the geometrical center of the associated SNR. Therefore one can
naturally explain why the tails behind a number of NSs [e.g. PSR
\object{B\,1757-24} (\cite{vgvaramadze-B2:fra94}; see also Sect. 3.2) or
a compact X-ray source in the SNR \object{IC\,433}
(\cite{vgvaramadze-B2:olb01})] do not point towards the centers
of their parent shells. In principle, the proper motion vector of a NS
even could be directed to the center of the SNR (just this situation takes
place in the case of  PSR \object{B\,0656+14}, which is located within
the SNR \object{Monogem Ring}; see \cite{vgvaramadze-B2:tho94}).
Therefore it is not impossible that a NS
born far from the center of the former wind-driven bubble (now the center of
the SNR) will reach it after a while (a possible example: the 24\,000 yr old
PSR \object{J\,1811-1925} located close to the center of the nearly
circular SNR \object{G\,11.2-0.3}; see \cite{vgvaramadze-B2:tor97}). From
this follows that the age of a NS inferred from the NS displacement
from the geometrical center of the associated SNR could be considerably
underestimated (cf. \cite{vgvaramadze-B2:kas01} and
\cite{vgvaramadze-B2:mig02}).

The fourth criterion should be applied for those claimed
associations, where the NS is located not far (at least in
projection) from the SNR's shell, e.g. in the case of PSR
\object{B\,1706-44}, which is superposed on the arc-like ``shell" of the SNR
\object{G\,343.1-2.3}. One of the arguments against the association between
these objects is the absence of any signs of interaction between
the pulsar and the SNR's shell (e.g. \cite{vgvaramadze-B2:nic96}).
But the apparent location of a NS on the edge of
SNR's shell simply could be due to the effect of projection in
nonspherically-symmetric SNRs (\cite{vgvaramadze-B2:boc02}).

\section{Some examples}

\subsection{PSR B~1610-50/SNR Kes~32}

Sometimes the high implied transverse velocities of NSs are used
to discard the possible NS/SNR associations. For example,
\cite*{vgvaramadze-B2:sta99} suggested (cf.
\cite{vgvaramadze-B2:piv00}) that the lack of a pulsar wind radio
nebula around  PSR \object{B\,1610-50} means that the maximum
space velocity of this pulsar is $450 d_5 \, {\rm km} \, {\rm
s}^{-1}$, where $d_5$ is the distance to the pulsar in units of 5
kpc, and therefore it could not be associated with the nearby SNR
\object{Kes\,32} since this association implies the transverse
velocity of the pulsar of $\simeq 2000 \, {\rm km}\,{\rm s}^{-1}$
(\cite{vgvaramadze-B2:car93}). The implied velocity, however,
could be reduced two times simply due to the possible off-centered
cavity SN explosion, and once again two or even more times if the
braking index of the pulsar is similar, respectively, to that of
PSR \object{B\,0540-69} ($n=1.8$) or the \object{Vela pulsar}
($n=1.4$).

\subsection{PSR B\,1757-24/SNR G\,5.4-1.2}

The high transverse velocity also was inferred for PSR
\object{B\,1757-24}, which lies well outside the shell of the SNR
\object{G\,5.4-1.2} (e.g. \cite{vgvaramadze-B2:cas87}).
The physical association of these two objects was
firmly established after the discovery
(e.g. \cite{vgvaramadze-B2:frai91}) of a tail of radio
emission connecting the pulsar with the SNR. However, the 
pulsar PSR
\object{B\,1757-24} is more interesting in that its proper
motion vector does not point away from the geometrical center
of the SNR (\cite{vgvaramadze-B2:fra94}).
We suggest that the SNR \object{G\,5.4-1.2} is the result of an
off-centered SN explosion in the preexisting wind-driven bubble
surrounded by a massive shell. The mass of the shell is a very
important parameter since it determines the evolution of the SNR:
if the mass of the shell is $\geq 50$ times the mass of the SN
ejecta, the SN blast wave merges with the shell (e.g.
\cite{vgvaramadze-B2:fra91}) and evolves into a momentum-conserving stage
(i.e. propagates with a velocity $< 200 \, {\rm km} \, {\rm s}^{-1}$).
In this case, even a young NS moving with a moderate
velocity ($\geq 200 \, {\rm km} \, {\rm s}^{-1}$) is able to
overrun the SNR's shell (cf. \cite{vgvaramadze-B2:gae95}),
provided that it was born not far from the edge of the wind-driven
bubble.
Our suggestion allows to reduce considerably the transverse velocity of the
pulsar\footnote{An indirect support to this suggestion comes from the
recent observations of the radio nebula surrounding PSR
\object{B\,1757-24} (\cite{vgvaramadze-B2:gae00}), which
showed that the proper motion velocity of
this pulsar should be much less than the implied one.} and naturally
explains why the tail behind the pulsar does not
point back to the center of the SNR.

\subsection{PSR B\,1706-44/SNR G\,343.1-2.3}

The reader is referred to \cite*{vgvaramadze-B2:boc02} who provide
some arguments in support of this association in details.

\section{Peculiar SNRs}

A concept of an off-centered cavity SN explosion could be used to
explain the peculiar structure of a number of SNRs and thereby to infer
the ``true" SN explosion sites in these SNRs. The later, in its turn,
could be used for searches for stellar remnants possible associated with
these SNRs.

\subsection{RCW\,86}

We suggest that the SNR \object{RCW\,86}
is the result of a cavity SN explosion of a
moving massive star (cf. \cite{vgvaramadze-B2:vin97}),
which ends its evolution just
near the edge of the main-sequence (MS) bubble. We also suggest that the
bright protrusion in the southwest half of the SNR
is the recently shocked dense material of a bow shock-like structure
generated by the moving SN progenitor star during the red supergiant
(RSG) phase [we interpret a clumpy optical arc of radius of $\simeq
1.5 d_{2.8}$ pc located interior to the X-ray and radio outlines of the
protrusion (see, e.g., Fig. 2 of \cite{vgvaramadze-B2:ros96}) as the
remnant of this structure], while the remainder of
the SNR is due to the interaction of the SN blast wave with the wall of
the adjacent MS bubble.
For the SNR's age of few times $10^3$ yr (i.e. the time required for the
SN blast wave to cross the MS bubble of diameter of $\simeq 35$ pc) and
provided that the NS was born with the velocity of $\simeq 200 \,
{\rm km} \, {\rm s}^{-1}$, one can propose that the stellar remnant
should be still within the region bounded by the optical arc.
To check this proposal we analysed the archival {\it ROSAT}
data, but the limited photon statistics and the moderate spatial
resolution of the {\it ROSAT} PSPC did not allowed the detection of a possible
compact X-ray source against the bright background emission of the SNR's
shell. The analysis of the {\it Chandra} data (which shortly
will be publicly available) would be highly desirable.

\subsection{1E 0102.2-7219}

A massive star moving with a sufficiently large velocity ($> {\rm few} \,
{\rm km} \, {\rm s}^{-1}$) crosses the MS bubble and during the RSG phase
moves mainly through the unperturbed interstellar medium. The slow, dense
wind losing during the RSG phase assumes the form of an elongated tail
stretched behind the moving star. This asymmetric and massive ($\simeq
10 M_{\odot}$) circumstellar structure could survive the passage of the
SN blast wave and will appear as a ``spoke" of thermal X-ray emission
joining the SN explosion site to the circular shell of the young
(few times $10^3$ yr) SNR. We suggest that just this
situation takes place in the case of the SNR \object{1E 0102.2-7219}
and interpret a prominent curl on the south end of the ``spoke"
(see Fig. 1 of \cite{vgvaramadze-B2:gaet00}) as the recently shocked wall
of the ``hollow" tail produced by the moving SN progenitor star.

In the above two examples we assumed that the SN explodes just after the RSG
phase, that is that the zero age MS mass of the SN progenitor star was
$\leq 20 M_{\odot}$ (e.g. \cite{vgvaramadze-B2:van98}). If, however, the
progenitor star is more massive, then after the RSG phase it evolves through
the Wolf-Rayet (WR) phase. The existence of this additional evolutionary
phase could have some important consequences. One of them is discussed below.

\section{Mixed-morphology SNRs}

We propose that the SN explosion sites in some middle-aged
(shell-like) SNRs could be marked by nebulae of thermal X-ray
emission and that this emission could be responsible for the
centrally-peaked X-ray appearance of some (so-called
mixed-morphology; \cite{vgvaramadze-B2:rho98}) SNRs. Our proposal
is based on the following simple arguments (see also
\cite{vgvaramadze-B2:gva02}).

A massive star, before it explodes as a SN, loses during the RSG
phase a considerable fraction ($\geq 2/3$) of its initial mass in
the form of a slow, dense wind, which occupies a region of radius
a few pc. The subsequent interaction of the fast WR wind with
the slow RSG wind results in the origin of dense clumps, radially
moving with velocities of $\geq 100\,{\rm km}\,{\rm s}^{-1}$ (see
\cite{vgvaramadze-B2:gva01} and references therein). After the SN
explodes, the SN blast wave propagates through the tenuous
interclump medium, leaving behind the dense clumps embedded in the
hot shocked interclump gas. The gradual evaporation of the
material of radially moving clumps results in the origin of an
expanding nebula of thermal X-ray emission, which is centered on the
SN explosion site\footnote{The centrally-filled X-ray appearance of
some SNRs could be connected with the development of large-scale
Rayleigh-Taylor deformations of the preexisting wind-driven
shell (induced by the impact of the SN blast wave), which results
in the increase of the effective ``thickness" of the SNR's shell.
For example, this effect could be responsible for the absence of
the limb-brightening in the northwest and southeast quadrants of
the \object{Vela SNR} (\cite{vgvaramadze-B2:gva99a}). For different
points of view on the origin of the mixed-morphology SNRs see
\cite*{vgvaramadze-B2:whi91} and \cite*{vgvaramadze-B2:pet01}.}.

Let us consider two mixed-morphology SNRs.

\subsection{3C\,400.2}

The SNR \object{3C\,400.2} consists of two circular radio
shells with the centrally-filled thermal X-ray emission peaked on the
region where the radio shells overlap each other (\cite{vgvaramadze-B2:dub94};
see also \cite{vgvaramadze-B2:yos01}).
We suggest that this SNR is the result of a SN explosion inside
the large-scale WR bubble adjacent to the MS bubble (cf.
\cite{vgvaramadze-B2:dub94}; \cite{vgvaramadze-B2:vel01}).
We also suggest that the (thermal) X-ray
emission comes from the hot gas evaporated from the dense circumstellar
clumps (see above) and that the X-ray peak coincides with the
SN explosion site. Note that the mass of X-ray emitting material
is $\simeq 10 M_{\odot} d_{2.3} ^{2.5}$ (\cite{vgvaramadze-B2:yos01}),
where $d_{2.3}$ is the distance to the SNR in units of 2.3 kpc
(\cite{vgvaramadze-B2:gia98}), whereas the expected mass of circumstellar
gas (i.e. the mass lost during the RSG phase) is $> 15 M_{\odot}$.

\subsection{G\,290.1-0.8}

The SNR \object{G\,290.1-0.8} consists of an elongated radio shell
and a central nebula of thermal X-ray emission. The unabsorbed 0.5-10.0
keV flux from the SNR is $1.8\times 10^{-10} \, {\rm erg} \,
{\rm cm}^{-2} \, {\rm s}^{-1}$ (\cite{vgvaramadze-B2:sla02}),
that at a distance of 7 kpc corresponds to a luminosity $L \simeq 10^{36}
d_7 ^2 \, {\rm erg} \, {\rm s}^{-1}$. We suggest that the SN explodes
far from the MS bubble, but inside the large-scale WR bubble (surrounded
by a massive shell; see below).
To estimate the mass of X-ray emitting gas, $M_{\rm X}$,
we assume that most of the observed X-ray flux comes from the
bright central nebula of radius $R \simeq 4 d_7$ pc.
Then assuming that the gas is homogeneously distributed over the whole volume
of the nebula, one has $M_{\rm X} = (4\pi R^3 L/3\Lambda
)^{1/2} \, m_{\rm H}$, where $\Lambda = 6.2\times 10^{-19} T^{-0.6} \,
{\rm erg} \, {\rm cm}^{-2}$ is the cooling function for temperatures between
$10^5$ K and $4\times 10^7$ K (e.g. \cite{vgvaramadze-B2:cow81}), and
$m_{\rm H}$ is the mass of a hydrogen atom.
For $T\simeq 7.2\times 10^6$ K (\cite{vgvaramadze-B2:sla02}),
one has $M_{\rm X} \simeq 10 d_7 ^{2.5}
M_{\odot}$, that is a quite reasonable value if the initial mass of the
SN progenitor star was $\geq 20 M_{\odot}$ (in this case the
progenitor star ends its evolution as a WR star).

We note that the size and the general appearance of the SNR
\object{G\,290.1-0.8}
[elongated shell, roughly parallel to the galactic plane; bright optical
emission on either side from the major axis of the SNR (e.g.
\cite{vgvaramadze-B2:ell79})] suggest that the SN explosion
occurs inside a cavity surrounded by a massive (barrel-like) shell
swept-up by the fast WR wind. The large-scale interstellar magnetic
field (at low latitudes parallel to the plane of the Galaxy)
accumulated in the wind-driven shell reduces the column density near the
magnetic poles of the shell (\cite{vgvaramadze-B2:fer91}), that leads to
the elongated shape of the resulting SNR and could be responsible for the
bilateral distribution of the optical emission along the SNR's shell
(the SN blast wave becomes radiative primarily near the magnetic equator,
where the column density is maximum; cf. \cite{vgvaramadze-B2:gva99b}).

\begin{acknowledgements}

I am grateful to W.Brinkmann for his support during my stay at the
Max-Planck-Institut f\"ur extraterrestrische Physik (Garching), where
this work was partially carried out. I am also grateful to R.Petre for his
interest to the paper and to the LOC for financial support.
This work was partially supported by the Deutscher Akademischer Austausch
Dienst (DAAD).

\end{acknowledgements}

\end{document}